\documentclass[a4paper,fleqn,usenatbib,usedcolumn,useAMD]{mnras}

\usepackage{amsmath}
\usepackage{amssymb}
\usepackage{txfonts}
\usepackage{tabularx}

\usepackage[T1]{fontenc}
\usepackage{ae,aecompl}

\usepackage{epsfig}

\title[Differential speckle polarimetry]{Differential speckle polarimetry at Cassegrain and Nasmyth foci}

\author[B. Safonov et al.]{
Boris Safonov,$^{1}$\thanks{E-mail: safonov@sai.msu.ru (BS)}
Pavel Lysenko,$^{2}$
Maria Goliguzova$^{2}$
and Dmitry Cheryasov$^{1}$
\\
$^{1}$Sternberg Astronomical Institute of M.V.Lomonosov Moscow State University, 119234, Moscow, Universitetskii pr-t, 13\\
$^{2}$Faculty of Physics of M.V.Lomonosov Moscow State University, 119991, Moscow, Leninskie Gory\\
}

\date{Accepted XXX. Received YYY; in original form ZZZ}

\pubyear{2018}

\begin{document}
\label{firstpage}
\pagerange{\pageref{firstpage}--\pageref{lastpage}}
\maketitle

\begin{abstract}
Polarimetric interferometry is a method allowing the study of the distribution of polarized flux at diffraction--limited resolution. Its basic observable is the ratio $\mathcal{R}$ of the visibilities of the object in two orthogonal polarizations. Here, we demonstrate how this observables can be measured with the SPeckle Polarimeter (SPP) of the 2.5-m telescope. The SPP is a combination of a dual--beam polarimeter and an EMCCD--based visible--range speckle interferometer. We propose a simple method for the correction of $\mathcal{R}$ for the instrumental polarization and polarization differential aberrations of the telescope. The polarized intensity image can be estimated from $\mathcal{R}$ under the assumption that the object is a point--like unpolarized source plus a faint extended polarized envelope. The phase of $\mathcal{R}$ can be used for measurement of the polaroastrometric signal --- the difference between the photocentres of orthogonally polarized images of the object. We investigate both possibilities using observations of unpolarized stars and stars with a significant polarized circumstellar environment --- $\mu$~Cep and RY~Tau.
\end{abstract}




\section{Introduction}

It is difficult to determine the origin of intrinsic polarization of some astrophysical objects. In the case of the UX Ori variables, polarization can be generated by either scattering on the circumstellar environment or extinction of stellar radiation on aligned dust \citep{Bastien1987,Li2016}. For some active galactic nuclei with strong polarimetric variability, polarization of radiation of the central machine \citep{Marscher2014} or polarization of the jet \citep{Marscher2010} occurs. For supernovae polarization can be the consequence of scattering on a nearby cloud or extinction on aligned dust along the line of sight \citep{Cikota2017}. 


High contrast is not necessary for the discrimination between these hypotheses, as long as it is known from conventional polarimetry that $1-10$ per cent of the radiation is polarized. However, it is more important to have the capability to study the polarized flux distribution at diffraction--limited resolution.

One of the methods allowing such diagnosis is a combination of polarimetry and stellar interferometry, including single--aperture interferometry. This technique has been demonstrated, e.g., by \citet{Norris2012} in an aperture--masking experiment using adaptive optics. They considered the ratio of the visibility of the object in horizontal and vertical polarizations as the basic observable of the method; a similar observable can be introduced for diagonal polarization. We will denote these ratios as $\mathcal{R}_Q$ and $\mathcal{R}_U$.

However adaptive optics is not strictly necessary if one wants to suppress the effect of atmospheric optical turbulence and reach the diffraction--limited resolution of a single--aperture telescope. There are still methods based on post--processing of short--exposure images, such as lucky imaging and speckle interferometry. These methods are not very demanding in terms of complex equipment and have experienced a resurgence in the last two decades thanks to the advent of Electron Multiplication CCD technology (EMCCD). They can be directly combined with polarimetry. The Extreme Polarimeter by \citet{Canovas2011} is a noteworthy example, with its methodology being similar to lucky imaging. 

Previously, \citet{Schertl2000} demonstrated NIR speckle interferometry in the two beams of a polarimeter. They called the method ``speckle polarimetry''. Their instrument simultaneously acquired a series of short--exposure images in two orthogonal polarizations. The instrument also featured a rotating half--wave plate, the position of which was fixed at $0^{\circ}$, $22.5^{\circ}$, $45^{\circ}$ or $67.5^{\circ}$ during acquisition of the series. The images corresponding to the two beams of the polarimeter were obtained from the series using bispectrum image--restoration techniques. Then, the polarimetric observables in image space were calculated.





 
From 2014--2015, we developed and constructed an instrument similar to that of \citet{Schertl2000} but for the visible range --- the SPeckle Polarimeter (SPP). The SPP is a combination of a dual--beam polarimeter with a rotating half--wave plate (HWP) and an EMCCD--based speckle interferometer. The difference in observational methodology compared with \citep{Schertl2000} is that we rotate the HWP continuously throughout the series acquisition thus more effectively suppress the optical turbulence effect. The instrument is installed at the Sternberg Astronomical Institute 2.5-m telescope \citep{Safonov2017}.

In the current paper, we demonstrate how differential polarimetric techniques can be applied in Fourier space and how to estimate $\mathcal{R}_Q$ and $\mathcal{R}_U$ values using the series obtained with SPP. The resulting values can be used directly for constraining the polarized flux of the object at the diffraction--limited resolution. We call the corresponding method differential speckle polarimetry (DSP) to emphasize the fact that we are comparing Fourier spectra in order to obtain polarimetric observables rather than computing the latter from reconstructed images (``speckle polarimetry'' according to \citet{Schertl2000}). Two approaches to the interpretation of $\mathcal{R}_Q$ and $\mathcal{R}_U$ values are discussed: polarimetric image restoration and so--called polaroastrometry \citep{Safonov2015}. 


The paper is organized as follows. In Section \ref{sec:method}, we describe the basics of polarimetric interferometry, instrument design and the processing of data. Section \ref{sec:instrpol} is dedicated to instrumental polarization effects. In the discussion in Section (\ref{sec:discussion}), the performance of DSP is presented and compared with other methods. The main outcomes of the work and further plans are given in Section \ref{sec:conclusion}. The appendices contain auxiliary results.


\section{Method}
\label{sec:method}

\subsection{Polarimetric interferometry}

The Stokes parameters are connected to the total intensity $I$, fraction $p$ and angle $\chi$ of polarization in the following way:
\begin{equation}
I = I,
\end{equation}
\begin{equation}
Q = p I \cos (2 \chi) \cos (2 \psi),
\end{equation}
\begin{equation}
U = p I \sin (2 \chi) \cos (2 \psi),
\end{equation}
\begin{equation}
V = p I \sin (2 \psi),
\end{equation}
where angle $\psi$ characterizes circular polarization, which will not be considered in this work, thus assuming $\psi=0$. The angle of polarization is measured in the counter--clockwise direction from the North. We will manipulate with the dimensionless Stokes parameters $q=Q/I$ and $u=U/I$ as well.

For the characterization of extended objects, the Stokes parameters can be considered as functions of the coordinates in the celestial sphere: $I(\boldsymbol{\alpha})$, $Q(\boldsymbol{\alpha})$, $U(\boldsymbol{\alpha})$. Here, $\boldsymbol{\alpha}=(\alpha_x,\alpha_y)$ is a two-dimensional vector of angular coordinates, counted from a certain point. The exact choice of this point does not influence the result. The $OX$ axis is directed to the North and the $OY$ axis --- to the East\footnote{This choice is due to the fact that in astronomy the angle of polarization is measured from the direction to the North, while in physics the reference direction is $OX$ axis}. In the following, we will extensively use the Fourier transforms of the distribution of Stokes parameters in the object: 
\begin{equation}
I(\boldsymbol{\alpha})\underset{FT}{\rightarrow}\widetilde{I}(\boldsymbol{f}),\,\,Q(\boldsymbol{\alpha})\underset{FT}{\rightarrow}\widetilde{Q}(\boldsymbol{f}),\,\,U(\boldsymbol{\alpha})\underset{FT}{\rightarrow}\widetilde{U}(\boldsymbol{f}),
\label{eq:Ivis}
\end{equation}
where $\boldsymbol{f}$ is the vector of spatial frequency. The quantity $\widetilde{I}(\boldsymbol{f})$ is widely known as visibility.

Let us consider the following ratios constructed from values (\ref{eq:Ivis}):
\begin{equation}
\mathcal{R}_{Q,0}(\boldsymbol{f}) = \frac{\widetilde{I}(\boldsymbol{f})+\widetilde{Q}(\boldsymbol{f})}{\widetilde{I}(\boldsymbol{f})-\widetilde{Q}(\boldsymbol{f})},
\label{eq:RQzero}
\end{equation}
\begin{equation}
\mathcal{R}_{U,0}(\boldsymbol{f}) = \frac{\widetilde{I}(\boldsymbol{f})+\widetilde{U}(\boldsymbol{f})}{\widetilde{I}(\boldsymbol{f})-\widetilde{U}(\boldsymbol{f})}.
\label{eq:RUzero}
\end{equation}
One can think of the first value as the ratio of visibility of the object in horizontal and vertical polarizations. The second value as a similar meaning, but for diagonal polarizations. The absolute values of these values were suggested as observables by \cite{Norris2012}. Here, we treat them as complex functions of spatial frequencies, containing both amplitude and phase.

\subsection{Polaroastrometry}
\label{sec:astrom}

The basic observable in polaroastrometry is the so--called polaroastrometric signal which constitutes two vectors \citep{Safonov2015}:
\begin{equation}
\boldsymbol{\Delta}_Q = \frac{1}{2}\Biggl[\frac{\int \bigl(I(\boldsymbol{\alpha})+Q(\boldsymbol{\alpha})\bigr)\boldsymbol{\alpha} d \boldsymbol{\alpha}}{\int \bigl(I(\boldsymbol{\alpha})+Q(\boldsymbol{\alpha})\bigr) d \boldsymbol{\alpha}} - \frac{\int \bigl(I(\boldsymbol{\alpha})-Q(\boldsymbol{\alpha})\bigr)\boldsymbol{\alpha} d \boldsymbol{\alpha}}{\int \bigl(I(\boldsymbol{\alpha})-Q(\boldsymbol{\alpha})\bigr) d \boldsymbol{\alpha}}\Biggr],
\end{equation}
\begin{equation}
\boldsymbol{\Delta}_U = \frac{1}{2}\Biggl[\frac{\int \bigl(I(\boldsymbol{\alpha})+U(\boldsymbol{\alpha})\bigr)\boldsymbol{\alpha} d \boldsymbol{\alpha}}{\int \bigl(I(\boldsymbol{\alpha})+U(\boldsymbol{\alpha})\bigr) d \boldsymbol{\alpha}} - \frac{\int \bigl(I(\boldsymbol{\alpha})-U(\boldsymbol{\alpha})\bigr)\boldsymbol{\alpha} d \boldsymbol{\alpha}}{\int \bigl(I(\boldsymbol{\alpha})-U(\boldsymbol{\alpha})\bigr) d \boldsymbol{\alpha}}\Biggr].
\end{equation}
In other words, the polaroastrometric signal $\boldsymbol{\Delta}_Q$ is the half--difference between the vectors of the photocentres of horizontally and vertically polarized images. The polaroastrometric signal $\boldsymbol{\Delta}_U$ has the same meaning, but for images polarized at $45^{\circ}$ and $135^{\circ}$. The components of polaroastrometric signal will be denoted: $\boldsymbol{\Delta}_Q = (s_q^{\star},t_q^{\star})$ and $\boldsymbol{\Delta}_U = (s_u^{\star},t_u^{\star})$. They have the dimension of angle. The component $s$ corresponds to axis $OX$ and component $t$ --- to $OY$.

$\mathcal{R}_{Q,0}(\boldsymbol{f})$ and $\mathcal{R}_{U,0}(\boldsymbol{f})$ can be simplified if observable frequencies are much smaller than frequency corresponding to the object size (the object is much smaller than diffraction--limited resolution of the optical system):
\begin{equation}
\mathcal{R}_{Q,0}(\boldsymbol{f}) = 1 + i4\pi(s^{\star}_q f_x + t^{\star}_q f_y),
\label{eq:IPSQ}
\end{equation}
\begin{equation}
\mathcal{R}_{U,0}(\boldsymbol{f}) = 1 + i4\pi(s^{\star}_u f_x + t^{\star}_u f_y).
\label{eq:IPSU}
\end{equation}
From these equations it follows that the polaroastrometric signal characterizes the slopes of the phase of $\mathcal{R}_Q(\boldsymbol{f})$ and $\mathcal{R}_U(\boldsymbol{f})$. 

\subsection{Instrument}

\begin{figure}
\begin{center}
\includegraphics[width=6cm]{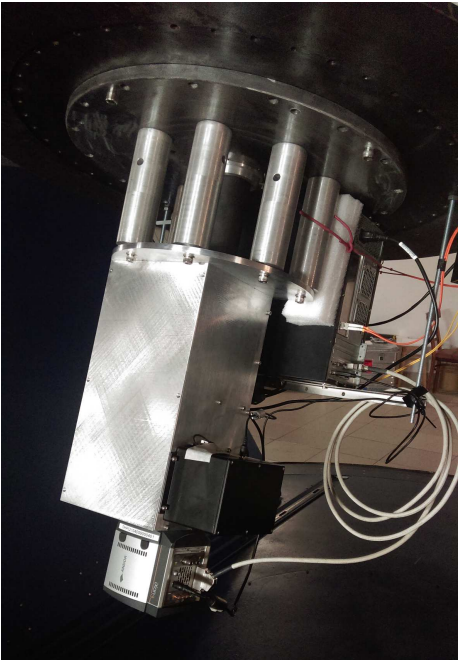}
\end{center}
\caption{SPeckle Polarimeter mounted at the Cassegrain focus of a 2.5-m telescope.
\label{fig:instrument}}
\end{figure}

In the current work, we use the data obtained with the SPeckle Polarimeter (SPP) of a 2.5-m telescope (see Figure~\ref{fig:instrument}). The SPP is a dual--beam polarimeter with a rotating half--wave plate (HWP). The Wollaston prism is used to split the beam into two orthogonally polarized beams that are subsequently used for the formation of two images on the same detector. The HWP rotates the polarization plane of incoming radiation in order to facilitate differential measurements and to detect all components of linear polarization. 

The instrument also employs an atmospheric dispersion compensator (ADC) and fast EMCCD as a detector. The effective angular scale of the instrument is 20.6 mas$\cdot$pix$^{-1}$. These features allow the estimation of Fourier spectra up to the cut--off frequency $D/\lambda$ for individual short-exposure images. This is characteristic of a speckle interferometer. We did not use aperture masks in our observations.

\begin{figure}
\begin{center}
\includegraphics[width=8cm]{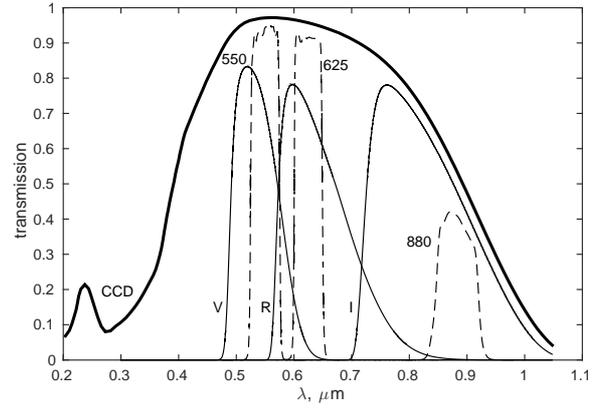}
\end{center}
\caption{Passbands of the filters used in SPP multiplied by the EMCCD quantum efficiency curve (thick black line).
\label{fig:passbands}}
\end{figure}

In SPP, 6 filters are available: standard $V$, $R_c$, $I_c$ Bessel filters and three interference medium band filters centred on 550, 625 and 880~nm. The former filters are used on faint targets, while the latter are more suitable for brighter objects. The passbands of the filters are presented in Figure~\ref{fig:passbands}. The design of the instrument and its basic calibration procedures were described in detail in a previous paper \citep{Safonov2017}.

During observations, the SPP was operated in the fast polarimetry regime. In this regime the detector continuously acquired the frames while the HWP rotated with constant angular speed. The period of the frames was 0.03~sec and the speed of the HWP was $300^{\circ}\cdot$sec$^{-1}$. Typically, the total number of accumulated frames was approximately $10^4$. 

The SPP can be mounted at either the Cassegrain or the Nasmyth focal station of the 2.5-m telescope. In the former case the whole system upstream of the HWP is axisymmetrical. In the latter case, there is one oblique reflection at the tertiary mirror with well--known properties; see Section \ref{sec:IPmodel}. We will consider observations conducted in both foci. More details on the observing circumstances will be given in later sections.


\subsection{Processing}

The processing starts with the standard procedures for the reduction of speckle interferometric data as described in \citep{Safonov2017}. For each frame the following actions are executed:
\begin{enumerate}
\item
We use only frames free of cosmic particles. Thanks to the small exposure time and small area of detector, only 0.1 per cent of the frames contain cosmic particles and are dismissed at this stage.
\item
The master bias frame is subtracted.
\item
We choose the areas of the frame free of source image and estimate the background from them. This background value is subtracted from all the frames.
\item
We zero the pixels with a level of signal less than three times the standard deviation of the readout noise. This procedure is valid for operations with electron multiplication higher than 50, as long as the pixels with a small signal cannot contain the photoelectrons.
\item
We take the regions of the frame corresponding to the two beams of the polarimeter. The square windows containing the source image are selected in both of them. The size of the window approximately equals twice the FWHM of the image.
\item
We use these subframes to calculate the Fourier spectra.
\item
The spectra of images are corrected for the Wollaston prism distortion and ADC prisms distortion.
\item 
For observations conducted at the Nasmyth focus, the rotation is applied to the spectra in order to compensate for the field rotation (we do not use the derotator at the Nasmyth focus). 
\end{enumerate}

\begin{figure*}
\begin{center}
\includegraphics[width=18cm]{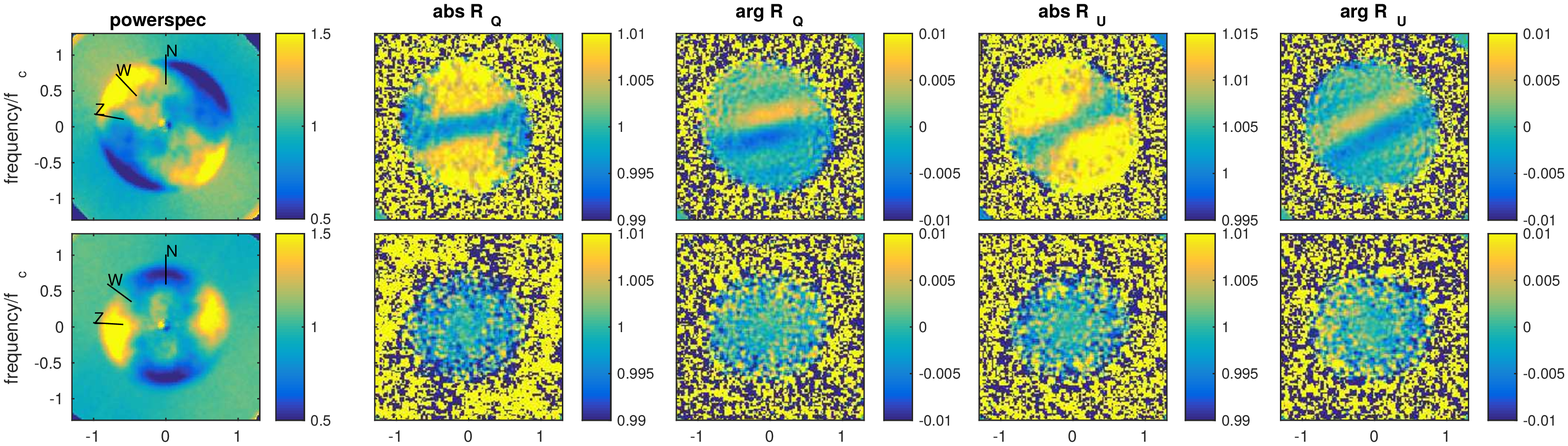}
\end{center}
\caption{Demonstration of speckle interferometric and DSP processing for the star $\mu$~Cep with a significant circumstellar scattering environment (top row) and an unpolarized star HIP109121 (bottom row). Both observations were made on 2017 December 2 at $\lambda=880$~nm. The leftmost column is the mean power spectrum of unpolarized images normalized by its azimuthal average. This kind of normalization is used for suppression of strong changes in the power spectrum in the radial direction \citep{Tokovinin2010}. The quadrupole pattern emerges due to residual atmospheric and Wollaston prism dispersion. The second to fifth columns stand for $|\mathcal{R}_Q|$, $\mathrm{arg} \mathcal{R}_Q$, $|\mathcal{R}_U|$, and $\mathrm{arg} \mathcal{R}_U$. In all panels, North is up and East is left. Some other important directions are shown in the leftmost panels: Z is the zenith and W is the Wollaston prism dispersion. The angular frequency normalized by the cut--off frequency $D/\lambda$ is along the axes. For details on observing circumstances see Table \ref{table:observationsNonpol}.
\label{fig:RvalPolnonpolCompar}}
\end{figure*}

In this way, we obtain spatial spectra of the two orthogonally polarized images formed by the Wollaston prism, i.e., $\widetilde{F}_{L,i}(\boldsymbol{f})$ and $\widetilde{F}_{R,i}(\boldsymbol{f})$, calling them left and right for the purposes of discussion.  For these spectra one can write:
\begin{equation}
\widetilde{F}_{L,i}(\boldsymbol{f}) = \mathrm{e}^{i2\pi(\rho_L\cdot\boldsymbol{f})}\bigl[\widetilde{I}_i(\boldsymbol{f}) + \widetilde{Q}_i(\boldsymbol{f}) \cos(4\theta_i) + \widetilde{U}_i(\boldsymbol{f}) \sin(4\theta_i)\bigr],
\end{equation}
\begin{equation}
\widetilde{F}_{R,i}(\boldsymbol{f}) = \mathrm{e}^{i2\pi(\rho_R\cdot\boldsymbol{f})}\bigl[\widetilde{I}_i(\boldsymbol{f}) - \widetilde{Q}_i(\boldsymbol{f}) \cos(4\theta_i) - \widetilde{U}_i(\boldsymbol{f}) \sin(4\theta_i)\bigr],
\end{equation}
where $\boldsymbol{f}$ is the vector of the two--dimensional spatial frequency, $i$ is the frame number, $\theta_i$ is the corresponding angle of HWP rotation, and $\widetilde{I}_i, \widetilde{Q}_i, \widetilde{U}_i$ are the quasi--instantaneous Fourier spectra of Stokes parameters distributions observed through the atmosphere and optical system of the telescope. 

The terms $\mathrm{e}^{i2\pi(\rho_L\cdot\boldsymbol{f})}$ and $\mathrm{e}^{i2\pi(\rho_R\cdot\boldsymbol{f})}$ take into account the unknown shifts $\rho_L$ and $\rho_R$ of the left and right beams of the polarimeter. For the subsequent processing, we need to know the difference between them. To determine this value we calculated the average $\langle\widetilde{F}_{L,i} \widetilde{F}_{R,i}^{*}\rangle$ and approximated its phase with a constant slope $\mathrm{e}^{i2\pi((\rho_R-\rho_L)\cdot\boldsymbol{f})}$. 

A set of $\widetilde{F}_{L,i}$ and $\widetilde{F}_{R,i}$ calculated for the series of frames is used for computation of values (the dependence on spatial frequency $\boldsymbol{f}$ is omitted for brevity):
\begin{equation}
\mathcal{R}_{ch} = 1+\frac{\bigl\langle (\widetilde{T} \widetilde{F}_{L,i} - \widetilde{F}_{R,i})  (\widetilde{T} \widetilde{F}_{L,i} + \widetilde{F}_{R,i})^*\cos(h \theta_i) \bigr\rangle}{\bigl\langle (\widetilde{T} \widetilde{F}_{L,i} + \widetilde{F}_{R,i})  (\widetilde{T} \widetilde{F}_{L,i} + \widetilde{F}_{R,i})^* \bigr\rangle - N_e^{-1}},
\label{eq:RavgC}
\end{equation}
\begin{equation}
\mathcal{R}_{sh} = 1+\frac{\bigl\langle (\widetilde{T} \widetilde{F}_{L,i} - \widetilde{F}_{R,i})  (\widetilde{T} \widetilde{F}_{L,i} + \widetilde{F}_{R,i})^*\sin(h \theta_i) \bigr\rangle}{\bigl\langle (\widetilde{T} \widetilde{F}_{L,i} + \widetilde{F}_{R,i})  (\widetilde{T} \widetilde{F}_{L,i} + \widetilde{F}_{R,i})^* \bigr\rangle - N_e^{-1}},
\label{eq:RavgS}
\end{equation}
where $h$ is the harmonics number, and $N_{e}$ is the average number of photons in a single frame (this term takes into account photon noise). The angle $\theta_i$ is estimated by introducing the linear polarizer with known orientation in the beam at the beginning and in the end of the series accumulation \citep{Safonov2017}. As long as the detector obtains $\approx40$ frames per revolution of the HWP, we calculated values (\ref{eq:RavgC}) and (\ref{eq:RavgS}) up to $N_h=20$.


The term $\widetilde{T}=\mathrm{e}^{i2\pi((\rho_R-\rho_L)\cdot\boldsymbol{f})}$ accounts for the difference in the displacements of the left and right beams of the polarimeter and is computed from the known value $\rho_L-\rho_R$ (see above).

Equations (\ref{eq:RavgC}) and (\ref{eq:RavgS}) are the adaptation of equation (6) from \citep{Safonov2013} for the instrument with a continuously rotating HWP. While the rotating HWP introduces the modulation, the post--processing in accordance with equations (\ref{eq:RavgC}) and (\ref{eq:RavgS}) represents a demodulation. 

$\mathcal{R}_{ch}$ and $\mathcal{R}_{sh}$ values computed for $h=4$ carry the useful signal. They correspond to Stokes $Q$ and $U$, respectively. We will denote them $\mathcal{R}_Q$ and $\mathcal{R}_U$. In Appendix \ref{app:averR} we show that $\mathcal{R}_Q$ and $\mathcal{R}_U$ are the estimations of visibility ratios introduced in (\ref{eq:RQzero}) and (\ref{eq:RUzero}):
\begin{equation}
\mathcal{R}_Q = \mathcal{R}_{Q,\mathrm{ins}}\mathcal{R}_{Q,0},
\label{eq:RobsQ}
\end{equation}
\begin{equation}
\mathcal{R}_U = \mathcal{R}_{U,\mathrm{ins}}\mathcal{R}_{U,0}.
\label{eq:RobsU}
\end{equation}
Here, $\mathcal{R}_{Q,\mathrm{ins}}$ and $\mathcal{R}_{U,\mathrm{ins}}$ are values depending on the optical scheme upstream of the HWP (modulator). We will discuss them in Section \ref{sec:instrpol}.



\begin{table*}
\begin{footnotesize}
\begin{center}
\caption{Polarimetric observations. UT, zenith distance $z$ and position angle P.A. are given at exposure centre. $\beta$ is the seeing measured with MASS-DIMM \citep{Kornilov2014}. $N_\mathrm{fr}$ is the number of accumulated photons, $N_\mathrm{ph}$ is number of accumulated photons. Ref. sys. is for reference system in which dimensionless Stokes parameters $q$ and $u$ are presented: equ means equatorial, hor means horizontal. \label{table:observationsNonpol}}
\tabcolsep=0.12cm
\begin{tabular}{cccccccccccc}
\hline
focus & object & UT        & z           & PA         & filter & $\beta$           & $N_\mathrm{fr}$ & $N_\mathrm{ph}$ & ref. sys. & $q\times10^4$ & $u\times10^4$ \\
      &        &           & $^{\circ}$  & $^{\circ}$ &      & arcsec &                 &         &         &               &  \\
\hline
C & $\mu$ Cep & 2017-12-02 & 42.7 & 315.5 & $880$ & 0.7 & 8527 & $1.1\times10^{11}$ & equ & $-25.2\pm5.8$ & $39.1\pm5.8$ \\ 
C & HIP109121 & 2017-12-02 & 17.4 & 305.2 & $880$ & 0.7 & 10319 & $1.8\times10^8$   & hor & $+13.8\pm5.9$ & $+9.5\pm5.9$  \\
C & RY Tau    & 2017-03-09 & 50.8 & 219.4 & $I$   & 1.0 & 8945 & $1.3\times10^8$    & equ & $+236.4\pm6.9$& $170.7\pm6.9$ \\
\hline
\end{tabular}
\end{center}
\end{footnotesize}
\end{table*}

\begin{table*}
\begin{footnotesize}
\begin{center}
\caption{Polaroastrometric measurements for observations presented in Table \ref{table:observationsNonpol}. Ref. sys. is for the reference system in which the polaroastrometric signal is presented: equ means equatorial (corrected for instrumental polarization effects), hor means horizontal (uncorrected for instrumental polarization effects). $s^{\star}_q$, $t^{\star}_q$, $s^{\star}_u$, and $t^{\star}_u$ are components of the polaroastrometric signal. $f_\mathrm{high}$ is the upper border of the area of $\mathcal{R}$ used for the computation of polaroastrometric signal. It is expressed in terms of the cut--off frequency $f_c$. $\chi_{r}^2$ is a reduced chi squared statistics for $\mathrm{arg}\mathcal{R}$ approximation by the plane. $\chi_{r,0}^2$ is computed under the assumption that $\mathcal{R}=1$. Tables \ref{table:observationsNonpol} and \ref{table:observationsNonpol2} are available in their entirety in a machine-readable format in the online journal.  A portion is shown in the text for guidance regarding its form and content.
\label{table:observationsNonpol2}}
\begin{tabular}{cccccccccccc}
\hline
focus & object & UT          & filter  & ref.sys. & $s^{\star}_q$   & $t^{\star}_q$ & $s^{\star}_u$ & $t^{\star}_u$ & $f_\mathrm{high}/f_c$ & $\chi^2_{r}$ & $\chi^2_{r,0}$   \\
      &        &           &   &       & \multicolumn{1}{c}{$\mu$as} & \multicolumn{1}{c}{$\mu$as} & \multicolumn{1}{c}{$\mu$as} & \multicolumn{1}{c}{$\mu$as} & & & \\
\hline
  C & $\mu$ Cep & 2017-12-02 & $880$ & equ & $+458\pm14$& $105\pm14$ & $188\pm14$ & $109\pm14$ & 0.13 & 6.8 & 114.6 \\ 
  C & HIP109121 & 2017-12-02 & $880$ & hor & $ +5\pm17$ & $+11\pm 23$ & $-20\pm17$ & $-2\pm 23$  & 0.70	& 1.1 & 1.1 \\
  C & RY Tau    & 2017-03-09 & $I$   & equ & $-1268\pm68$ & $-400\pm66$ & $1365\pm68$ & $593\pm66$ &  0.20 & 2.1 & 161.9 \\
\hline
\end{tabular}
\end{center}
\end{footnotesize}
\end{table*}

One can follow the analogy between this approach and the double--ratio method of polarimetry considered by \citet{Bagnulo2009} and \citet{Canovas2011}. The double--ratio is preferable over the double--difference here because it is used to get rid of atmospheric noise which has a multiplicative nature in Fourier space.

\begin{figure}
\begin{center}
\includegraphics[width=7.5cm]{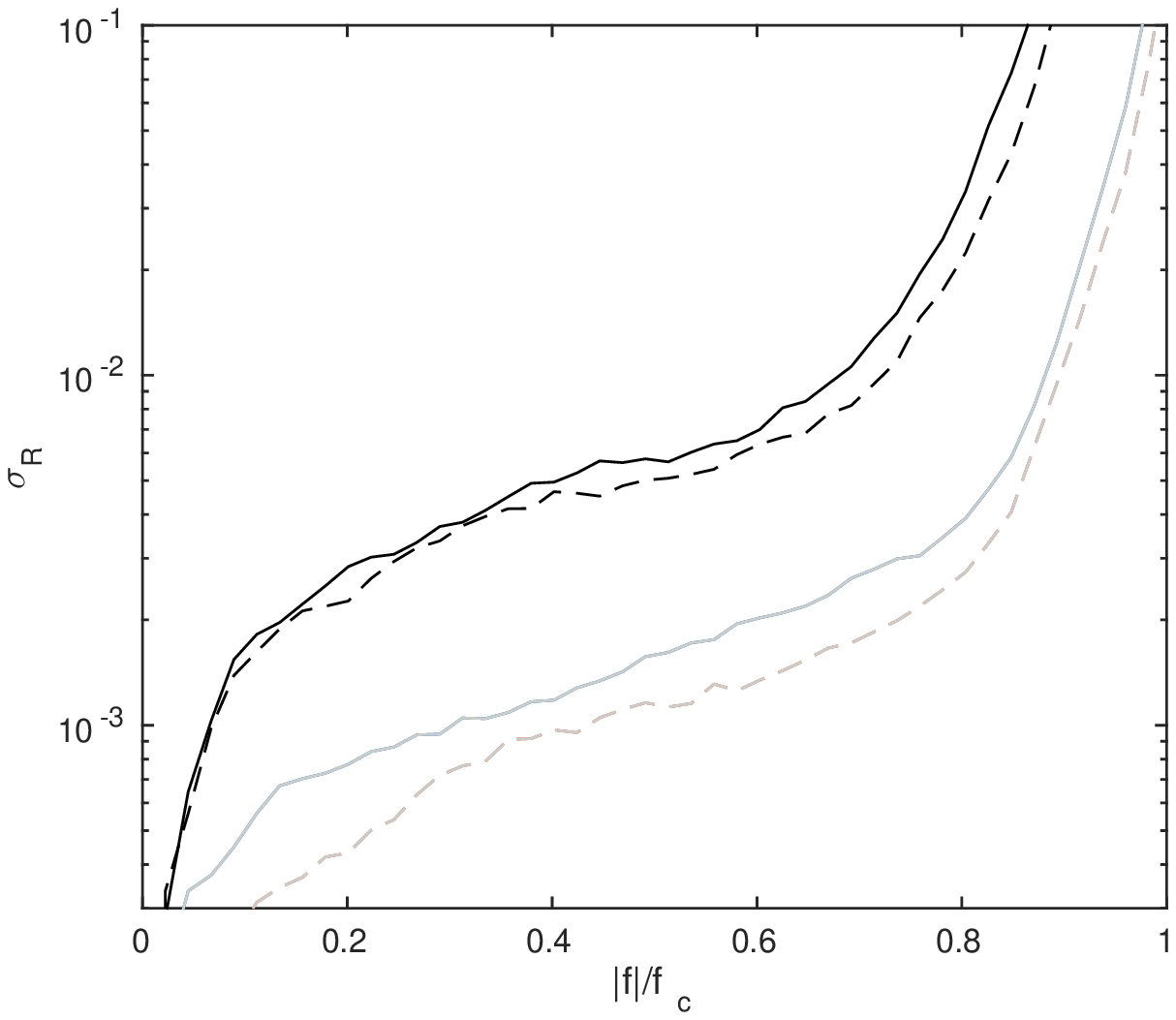}
\end{center}
\caption{Error of estimation of $\mathcal{R}$. Black lines are for HIP109121 observation, grey are for $\mu$~Cep. Both observations were made on 2017 December 2 at $\lambda=880$~nm (observing circumstances are provided in Table \ref{table:observationsNonpol}). Solid lines are for $\sigma_{\mathcal{R},\mathrm{abs}}$, and dashed lines are for $\sigma_{\mathcal{R},\mathrm{arg}}$.
\label{fig:Rnoise}}
\end{figure}

\begin{figure*}
\begin{center}
\includegraphics[width=18cm]{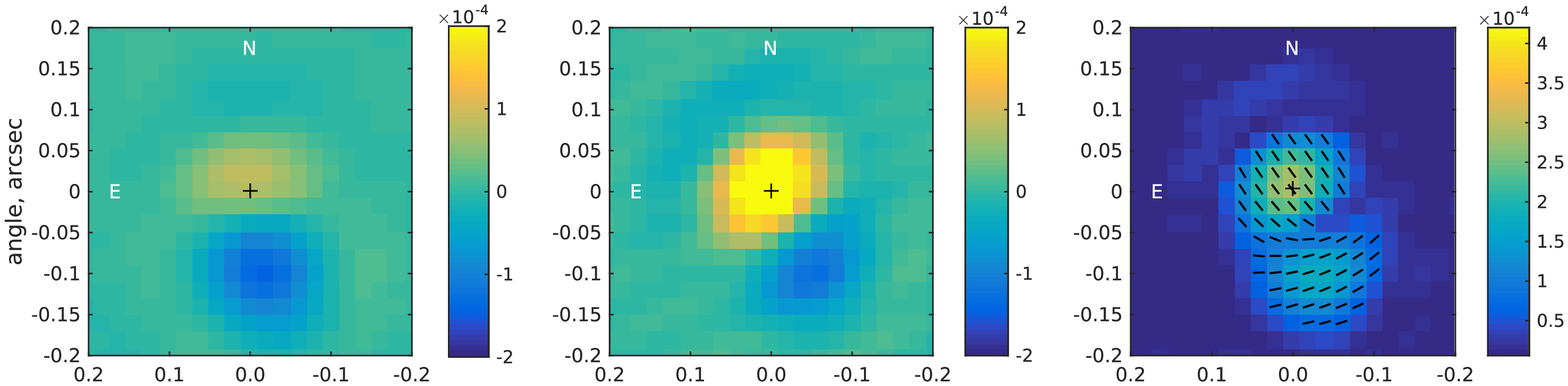}
\end{center}
\caption{The polarized flux in the vicinity of $\mu$~Cep is reconstructed by the method described in Section \ref{sec:recon}. The observation was made on 2017 December 2 at $\lambda=880$~nm. From left to right: Stokes $Q$, Stokes $U$, and polarized intensity $\sqrt{Q^2+U^2}$. In the latter panel the orientation of polarization is indicated by short black lines. The length of lines is chosen arbitrarily and does not reflect the polarization fraction as usual. The brightness scale is indicated by colour bars on the right side of panels. Units are the ratio of the polarized flux in the pixel to the total flux of the object (circumstellar envelope plus star). In all panels, North is up and East is left. The cross at coordinates (0, 0) stands for the position of the star. The unpolarized flux of the star is not visible. 
\label{fig:muCepImage}}
\end{figure*}

We do not claim that the proposed method of estimating $\mathcal{R}_Q$ and $\mathcal{R}_U$ is optimal. Variations of equations (\ref{eq:RavgC}) and (\ref{eq:RavgS}) that are more effective in terms of SNR in the resulting $\mathcal{R}$ likely exist.

The processing described is implemented using the Matlab programming language as a suite of methods\footnote{http://lnfm1.sai.msu.ru/kgo/mfcproc.tar.gz}. In these methods, we made use of the astronomy \& astrophysics toolbox\footnote{https://webhome.weizmann.ac.il/home/eofek/matlab/index.html} for Matlab by \citet{Ofek2014}. For the organization of processing, we employ PostgreSQL database where the information on raw observations is stored. 

\subsection{Example of $\mathcal{R}$ measurement}

In Figure \ref{fig:RvalPolnonpolCompar} the values of $\mathcal{R}$ for HIP109121 and $\mu$ Cep are presented and the respective observing circumstances are given in Table \ref{table:observationsNonpol}. One can see that for HIP109121, the amplitude and phase do not deviate significantly from unity and zero, respectively ($\chi_{r,0}^2=1.1$; for the definition see (\ref{eq:chir2def})). This is quite expected as long as HIP109121 is a single, relatively old main sequence star. It hardly possesses any dusty circumstellar environment detectable with SPP. Even if some part of the radiation is scattered in the stellar atmosphere, its effect on $\mathcal{R}$ will be negligible due to the small expected angular size of the star ($\approx0.18$~mas).

On the other hand, for $\mu$ Cep both the amplitude and phase of $\mathcal{R}$ are essentially non-zero. This indicates that the polarized flux distribution in this object differs from the total flux distribution. In the following subsections we discuss two possible approaches to interpretation of these values. 

In Appendix \ref{app:rotation} we give the methods of transformation of $\mathcal{R}$ in a rotated reference system and estimation of its error $\sigma_{\mathcal{R}}$. The dependence of $\sigma_{\mathcal{R}}$ on $|f|$ is shown in Figure \ref{fig:Rnoise}. For HIP109121 ($I=6.0$) it equals $\approx5\times10^{-3}$ at $0.5 f_c$, where $f_c$ is the cut--off frequency $D/\lambda$. This level of noise is comparable with the values obtained by \citet{Norris2015}: $4.2\times10^{-3}$. Note that for the much brighter object $\mu$~Cep ($I=0.8$), the error is several times smaller. 


\subsection{$Q$ and $U$ Stokes distribution reconstruction}
\label{sec:recon}

Quite frequently an object possessing a polarized structure can be approximated by the model of an unpolarized unresolved star and a much--fainter polarized circumstellar envelope. For example, one can consider protoplanetary disks around pre--main sequence stars or dusty atmospheres of evolved stars. In both cases, the circumstellar environment contains dust that scatters stellar radiation and therefore generates polarized flux.

For the objects of this kind the distributions of Stokes parameters will be:
\begin{equation}
I(\boldsymbol{\alpha}) = I_\star(\boldsymbol{\alpha}) + I_c(\boldsymbol{\alpha}),\,\,\,Q(\boldsymbol{\alpha}) = Q_c(\boldsymbol{\alpha}),\,\,\,U(\boldsymbol{\alpha}) = U_c(\boldsymbol{\alpha}).
\label{eq:envStokes}
\end{equation}
Here, $I_\star$ is the intensity distribution in the stellar image. $I_c, Q_c, U_c$ are the distribution of Stokes parameters of the circumstellar cloud (all values are normalized by the total flux of the object). After the substitution of values (\ref{eq:envStokes}) into equations (\ref{eq:RQzero}) and (\ref{eq:RUzero}) we obtain:
\begin{equation}
\mathcal{R}_{Q,0} = \frac{1 - \Delta\widetilde{I}_\star + \widetilde{I}_c + \widetilde{Q}_c}{1 - \Delta\widetilde{I}_\star + \widetilde{I}_c - \widetilde{Q}_c},
\end{equation}
\begin{equation}
\mathcal{R}_{U,0} = \frac{1 - \Delta\widetilde{I}_\star + \widetilde{I}_c + \widetilde{U}_c}{1 - \Delta\widetilde{I}_\star + \widetilde{I}_c - \widetilde{U}_c}.
\end{equation}
Here, $\Delta\widetilde{I}_\star$ is the deviation in the visibility of the star from unity. $\Delta\widetilde{I}_\star\ll1$ because the star is unresolved. Taking into account the faintness of the envelope, it is possible to obtain the following expressions for $\mathcal{R}_{Q,0}$ and $\mathcal{R}_{U,0}$:
\begin{equation}
\mathcal{R}_{Q,0} \approx 1 + \frac{2\widetilde{Q}_c}{1-\Delta\widetilde{I}_\star + \widetilde{I}_c} \approx 1 + 2 \widetilde{Q}_c,
\end{equation}
\begin{equation}
\mathcal{R}_{U,0} \approx 1 + \frac{2\widetilde{U}_c}{1-\Delta\widetilde{I}_\star + \widetilde{I}_c} \approx 1 + 2 \widetilde{U}_c.
\end{equation}
Therefore, rough estimations of the $Q$ and $U$ Stokes distributions can be obtained from measurements of $\mathcal{R}_{Q,0} - 1$ and $\mathcal{R}_{U,0} - 1$ by applying inverse Fourier transform. From the above equations, it is clear that the relative precision of this estimation is on the order of the ratio of the flux of the envelope to the total flux of the object. Additional error can be introduced if the star is partially resolved. This error can be evaluated as the maximum decrease in visibility of the unresolved source at the longest baseline of the experiment.


We demonstrate this approach using the $\mu$~Cep data presented in the previous section. Before taking the inverse Fourier transform we multiplied the observational $\widetilde{Q}$ and $\widetilde{U}$ by the diffraction OTF of the telescope. The result is presented in Figure \ref{fig:muCepImage} as the distribution of Stokes $Q$ and $U$. The total polarized intensity and the polarization angle were calculated from these results:
\begin{equation}
PI = \sqrt{Q^2+U^2},\,\,\,\chi = \frac{1}{2}\mathrm{atan}(\frac{U}{Q}).
\end{equation}
These values are displayed in the right panel of Figure \ref{fig:muCepImage}. The ratio of the integral polarized flux to the total flux of the object is 0.035.

From observations of $\mu$~Cep with the Mark III Stellar Interferometer at 800~nm, it is known that the star has an angular diameter equal to $18.6\pm0.4$~mas \citep{Mozurkewich2003}. The corresponding fall in visibility at a 2.5~m baseline is 0.92. This sets the lower limit on the relative error of the determination of Stokes parameters at 0.08. 

On the other hand, \citet{Mozurkewich2003} report the presence of a fully resolved source containing 20 per cent of the flux from the object. The minimum baseline of the Mark III interferometer was 3~m, which corresponds to the angular scale of $\lambda/D=55$~mas. In our case, the accessible angular scales range is from 55~mas to 2.5 arcsec.

Thus, the polarized nebula presented in Figure \ref{fig:muCepImage} with an angular size of 0.15~arcsec may be an unresolved component discovered by \citet{Mozurkewich2003}. This is supported by the fact that the total polarized flux we found is roughly consistent with the fraction of unresolved unpolarized flux measured by \citet{Mozurkewich2003}. However, a direct comparison is impossible, since the average fraction of polarization in the envelope is not known. In addition, the difference in epochs is approximately 28 years, while the characteristic timescale of object variability is much less.

As long as the fraction of the envelope flux is 0.1--0.2 in this case, the relative precision of the determination of Stokes parameters is roughly the same. Therefore, the image in Figure \ref{fig:muCepImage} can be interpreted only qualitatively. A quantitative comparison of observations and models should be done in terms of $\mathcal{R}$ values. It is worth noting that the relative precision of the distribution of Stokes parameters  improves for fainter envelopes and for stars with smaller angular size.


Taking into account these reservations, we can draw the following cautious conclusions regarding the $\mu$ Cep envelope. There is a polarized nebulosity at a stellocentric distance of $0.1-0.15$~arcsec having PA=$170-220^{\circ}$. The polarization orientation in this nebulosity follows a so--called azimuthal polarization pattern; thus, it is probably a reflection nebula associated with the star. In addition, there is a polarized source superimposed on the star and having $\chi=40^{\circ}$. This is likely to be the cloud of the circumstellar envelope of $\mu$~Cep as well. 

The departure from central symmetry in the envelope of $\mu$~Cep was found before by e.g., \citet{deWit2008} in mid-IR on angular scales of $1-2$~arcsec. Also \citet{Harper2018} recently reported an asymmetry of mass loss at effective temperatures of $\approx550$~K using mid-IR spectroscopy. These observations along with ours are evidence for the inhomogeneity of mass loss.


\subsection{Extraction of polaroastrometric signal}
\label{sec:PSextraction}

In the case of $\mu$~Cep, one can clearly see stripes in $\mathcal{R}$ because the typical angular size of the polarized flux features is $>\lambda/D$. Sometimes the object can be smaller or much smaller than $\lambda/D$. In these cases the region of frequencies reflecting the structure will be much larger than the cut--off frequency. Nevertheless, the phase of $\mathcal{R}$ will preserve the slope near the origin of coordinates, which characterizes the polaroastrometric signal.

The polaroastrometric signal was obtained by fitting of the observational $\mathrm{arg}\mathcal{R}_{c4}(\boldsymbol{f})$ and $\mathrm{arg}\mathcal{R}_{s4}(\boldsymbol{f})$ with the planes (\ref{eq:IPSQ}) and (\ref{eq:IPSU}). The dependencies $\mathrm{arg}\mathcal{R}(\boldsymbol{f})$, corresponding to other harmonics were approximated in a similar way and were used in noise estimation, as described by \citet{Safonov2015}. The fitting was performed by the weighted least squares method in a certain domain of spatial frequencies $f_\mathrm{low}<|f|<f_\mathrm{high}$. The weights were set equal to the inverse squared error of $\mathcal{R}$ (see Appendix \ref{app:rotation}). For characterization of goodness of fit we averaged the reduced $\chi_r^2$ for $q$ and $u$ components.

The lower border of the approximation area was set as the frequency corresponding to a long--exposure image FWHM: $f_\mathrm{low}=\beta^{-1}$. Taking into account the signal on frequencies $f<f_\mathrm{low}$ leads to a significant change in the result and {an increase in} $\chi_r^2$. It is probably related to underestimation of the $\mathcal{R}$ noise at these frequencies. For the higher border of approximation $f_\mathrm{high}$ we took the frequency at which the SNR falls below 1. In an ideal case the $f_\mathrm{high}$ will be only slightly smaller than the cut--off frequency $f_c$. However, in practice, it is much smaller due to effects of finite exposure time, finite spectral bandwidth, photon noise and Wollaston prism dispersion. Both borders were determined by visual inspection.

\begin{figure*}
\includegraphics[width=18cm]{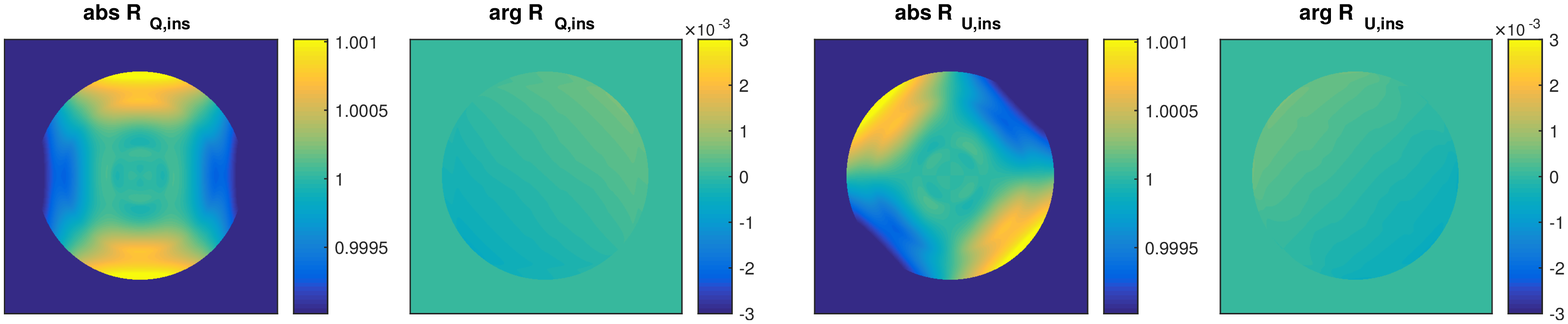}
\includegraphics[width=18cm]{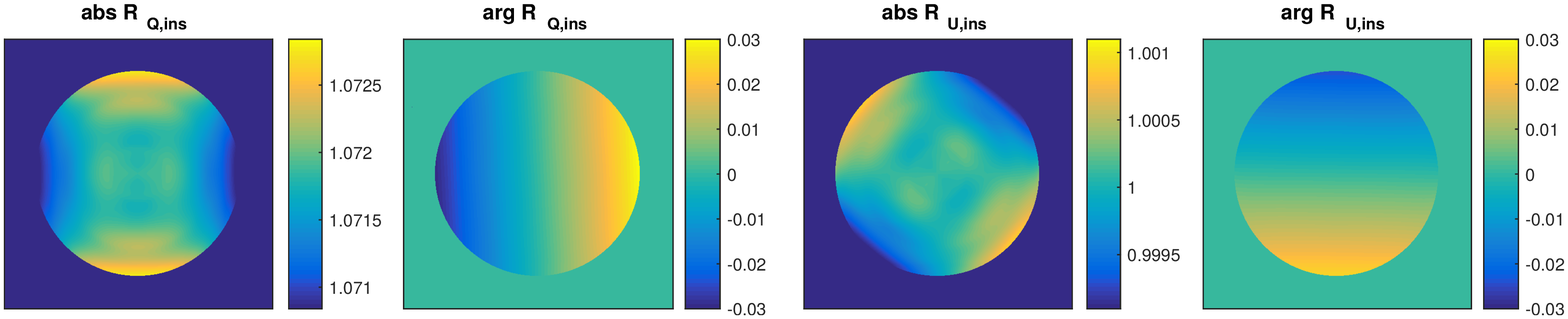}
\caption{Instrumental effect $\mathcal{R}_\mathrm{ins}$ calculated for the 2.5-m telescope at $\lambda=580$~nm. The top row corresponds to the Cassegrain focus and the bottom row to the Nasmyth focus. Size of displayed region of Fourier space corresponds to discretization with a period of 20.6 mas$\cdot$pix$^{-1}$ (angular scale of the SPP). The circle corresponds to the cut--off frequency.
\label{fig:Rexample}}
\end{figure*}

The approximation gives the polaroastrometric signal in the reference frame connected to the instrument. Its transformation to equatorial and horizontal reference frames is performed using formulae from \citep{Safonov2015}.

The polaroastrometric signal extracted from the series for HIP109121 and $\mu$~Cep is presented in Table \ref{table:observationsNonpol2}. One can see that for a main sequence star it is small but still significant for $\mu$~Cep. Note that we used only a small region of Fourier space for the determination of the signal in the case of $\mu$~Cep: $f<0.13f_c$. In a larger portion of Fourier space, the departure of $\mathrm{arg}\mathcal{R}$ from the plane becomes too obvious, even in the considered region $\chi_r^2=6.8$. This emphasizes again that polaroastrometry is applicable when the characteristic angular extent of the source $\lesssim\lambda/D$.

The typical precision of polaroastrometric signal measurement is $20$ $\mu$as for a $V=6$~mag star. A thorough study of precision will be presented in a separate paper.

\section{The effect of instrumental polarization}
\label{sec:instrpol}

\subsection{Model}
\label{sec:IPmodel}

During observations at the Nasmyth focus, the oblique reflection of the beam at the diagonal mirror leads to emergence of so--called instrumental polarization. For example, measurements of unpolarized stars show significant polarization, while measurements of polarized objects are modified. The change in polarization state by oblique reflection can be described by the formalism of Stokes vectors and Mueller matrices. We applied this approach to SPP \citep{Safonov2017}. In this previous work, on the basis of observations of unpolarized stars, we constructed a model of diagonal mirror coating: an aluminium reflective layer and protective layer of silicon dioxide 0.211 $\mu$m thick. The refractive index and absorption of aluminium were taken from \citet{McPeak2015} and multiplied by 0.91. Then, we computed the respective Mueller matrices for the wavelength range from 380 to 2200~nm. The Mueller matrix for a certain observation is computed by weighting them by the effective spectrum of detected radiation. Reversing the transformation described by this matrix, it is possible to remove the effect of instrumental polarization.

With the model of diagonal mirror coating and the model of the optical scheme of the telescope at hand, we calculated $\mathcal{R}_{Q,\mathrm{ins}}$ and $\mathcal{R}_{U,\mathrm{ins}}$ in the wavelength range from 440~nm to 1040~nm using the method presented in Appendix \ref{app:averR} and Zemax software. Instrumental polarization effects are associated with the telescope, which has an alt-az mount in our case. Therefore, hereinafter in this section, we will work in the horizontal reference system.

The example of $\mathcal{R}_{Q,\mathrm{ins}}$, $\mathcal{R}_{U,\mathrm{ins}}$ is presented in Figure \ref{fig:Rexample}. As one can see, an average of the absolute part of $\mathcal{R}_{Q,\mathrm{ins}}$ is 1.072, while the amplitude of the fluctuations is $\approx10^{-3}$. The $|\mathcal{R}_{U,\mathrm{ins}}|$ behaves similarly with an average value equal to unity. This is a direct consequence of the fact that at $\lambda=580$~nm instrumental polarization is directed along nadir--zenith line. At the same time, the variation in the phase of $\mathcal{R}_\mathrm{ins}$ is 30 times larger for both Stokes $Q$ and $U$ and manifests itself as a constant slope of phase. These slopes are produced by differential polarization aberrations caused by oblique reflection \citep{Breckinridge2015}. The effect was analysed in detail by \citet{Schmid2018} for the case of SPHERE/ZIMPOL.  They dubbed it ``polarimetric differential beamshift''.

All in all, $\mathcal{R}_\mathrm{ins}$ can be described by the following equations with acceptable accuracy ($10^{-3}$):
\begin{equation}
\mathcal{R}_{Q,\mathrm{ins}}(\boldsymbol{f}) = 1+2q_\mathrm{ins}+i4\pi(s^{\star}_{q,\mathrm{ins}} f_x + t^{\star}_{q,\mathrm{ins}} f_y), 
\label{eq:insSlope1}
\end{equation}
\begin{equation}
\mathcal{R}_{U,\mathrm{ins}}(\boldsymbol{f}) = 1+i4\pi(s^{\star}_{u,\mathrm{ins}} f_x + t^{\star}_{u,\mathrm{ins}} f_y).
\label{eq:insSlope2}
\end{equation}
Here, $q_\mathrm{ins}$ is the instrumental polarization, i.e., the fraction of polarization of an unpolarized star observed through the telescope \citep{Safonov2017}.

\begin{figure}
\begin{center}
\includegraphics[width=8cm]{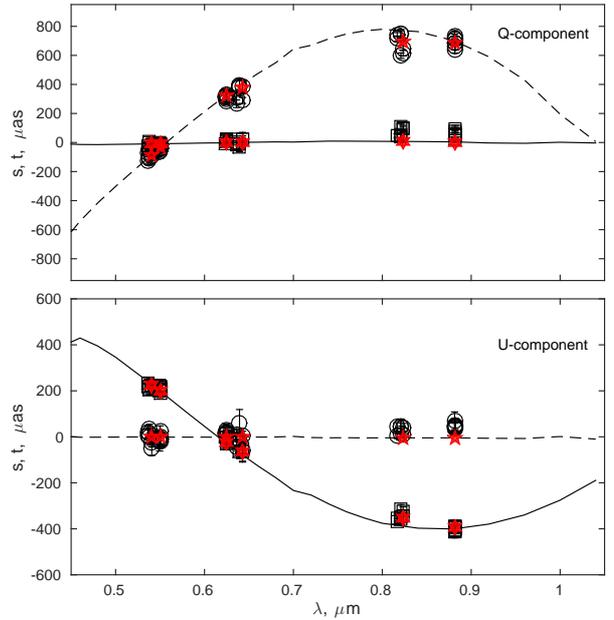}
\end{center}
\caption{Dependence of IPS on wavelength. Lines indicate the expected IPS, with solid and dashed lines corresponding to $s$ and $t$ components, respectively. The observed IPS is given by signs with error bars and the respective $X$--coordinate is determined as the effective wavelength for a given observation. Circles and squares correspond to $s$ and $t$ components, respectively. For details on observing circumstances see electronic version of Table \ref{table:observationsNonpol}. Red signs denote the expected IPS for a G3V star, pentagrams are $s$ components and hexagrams are $t$ components. 
\label{fig:instrPAsignal}}
\end{figure}

Comparing these equations with (\ref{eq:IPSQ}) and (\ref{eq:IPSU}), one can note that the phase of $\mathcal{R}_\mathrm{ins}$ can be interpreted in terms of the polaroastrometric signal (Sections \ref{sec:astrom} and \ref{sec:PSextraction}). We will call values $s^{\star}_{q,\mathrm{ins}}, t^{\star}_{q,\mathrm{ins}}, s^{\star}_{u,\mathrm{ins}}, t^{\star}_{u,\mathrm{ins}}$ the instrumental polaroastrometric signal (IPS). In Figure \ref{fig:instrPAsignal} the expected dependencies of IPS on wavelength are given. As one can see, while the components $s^{\star}_{q,\mathrm{ins}}$ and $t^{\star}_{u,\mathrm{ins}}$ are almost zero $\pm10~\mu$as, the amplitude of change of the other two components is $\approx800~\mu$as. 

In Figure \ref{fig:Rexample}, we give model $\mathcal{R}_\mathrm{ins}$ values computed for the Cassegrain focus for comparison. As expected for axisymmetrical configuration, they do not deviate by more than $10^{-3}$ from unity. The expected IPS for the Cassegrain focus is zero as well.

\begin{table}
\renewcommand{\thetable}{\arabic{table}}
\begin{center}
\caption{Stars that were used for estimation of IPS at the Cassegrain and Nasmyth foci. An undetectable level of intrinsic polarization and polaroastrometric signal is expected for them. Sp is spectral class. $l$ is galactic latitude. \label{table:nonpol}}
\begin{tabular}{ccccc}
\hline
star     & $V$  & Sp.  & distance & $l$ \\
         &      &      & pc   & $^{\circ}$\\
\hline
HIP3641   & 6.39 & F3V & 40.7 & +6.5   \\
HIP38325 & 6.02 & F6V  & 29.6 & +30.2 \\ 
HIP20156 & 5.40 & A7IV & 74.5 & -00.2 \\
HIP25143 & 5.54 & A3V  & 89.7 & +2.6  \\
HIP46125 & 6.68 & F0V  & 119  & +45.6 \\
HIP46873 & 6.88 & A4V  & 92.0 & +46.2 \\
HIP46963 & 6.54 & F0V  & 99.5 & +49.9 \\
HIP81800  & 6.46 & F8V & 29.5 & +40.9  \\
HIP89474  & 6.30 & G1V & 23.3 & +24.9  \\
HIP96901  & 6.20 & G3V & 21.2 & +13.2  \\
HIP100907 & 5.62 & A3V & 81.4 & -00.0  \\
HIP109121 & 6.18 & A3V & 97.8 & -08.4  \\
HIP116083 & 8.69 & F7V & 160  & -72.1  \\
\hline
\end{tabular}
\end{center}
\end{table}

\begin{table*}
\begin{footnotesize}
\begin{center}
\caption{
Average instrumental polarization and instrumental polaroastrometric signal. Exp. means expected from model, and obs. means observed. C is the Cassegrain focus, and N is the Nasmyth focus. AV0 and M2V are calculations for the typical spectra of A0V and M2V stars, respectively. Numbers after the $\pm$ sign mean the spread of observed values.
\label{table:IPAS}}
\tabcolsep=0.12cm
\newcolumntype{d}[1]{D{,}{\pm}{#1}}
\begin{tabular}{cccd{3}cd{3}cd{3}ccd{3}ccd{3}cd{3}}
\hline
\hline
focus & filter & \multicolumn{2}{c}{$q\times10^4$}     & \multicolumn{2}{c}{$u\times10^4$} & \multicolumn{2}{c}{$s_q^{\star}$, } &  \multicolumn{3}{c}{$t_q^{\star}$, } & \multicolumn{3}{c}{$s_u^{\star}$, } &  \multicolumn{2}{c}{$t_u^{\star}$, } \\
      &      & exp. & \multicolumn{1}{c}{obs.} & exp. & \multicolumn{1}{c}{obs.}& exp. & \multicolumn{1}{c}{obs.}  & \multicolumn{2}{c}{exp.} & \multicolumn{1}{c}{obs.} & \multicolumn{2}{c}{exp.}  & \multicolumn{1}{c}{obs.} &  exp. &\multicolumn{1}{c}{obs.} \\
      &      & A0V  &                          & A0V  &                         & A0V  &                           &      A0V &      M2V      &                          &      A0V &      M2V       &                          & A0V  &                         \\
  &        &       &            &   &          & $\mu$as & \multicolumn{1}{c}{$\mu$as} & $\mu$as & $\mu$as & \multicolumn{1}{c}{$\mu$as} & $\mu$as & $\mu$as & \multicolumn{1}{c}{$\mu$as} & $\mu$as & \multicolumn{1}{c}{$\mu$as}    \\
\hline
C & $V$      & 0     & 0.4,1.1    & 0 & -0.3,0.6 & 0    &   -1,14    & 0      & 0      &   8,17   & 0    & 0    & -13,16    & 0    &  4,13   \\
N & 550      & 338.3 & 344.2,4.9  & 0 & -1.4,2.5 & $-8$ &  -16,12    & $ -40$ & $-25$ &  -43,17  & 200  & 192  & +210,12   & $-1$ &  -10,15  \\ 
N & 625      & 361.5 & 385.0,2.4  & 0 & -1.2,3.0 & $-1$ &  +13,11    & $ 322$ & $335$ & +320,12  & -25  & -34  &  -19,11   & $-1$ &  +19,7   \\ 
N & 880      & 198.4 & 230.7,6.6  & 0 &  3.7,2.2 & $3$  &  +56,20    & $ 687$ & $684$ & +702,36  & -394 & -394 & -401,12   & $-5$ &  +52,12  \\ 
N & $V$      & 337.2 & 327.1,6.1  & 0 & -0.1,3.0 & $-9$ &  -18,17    & $-111$ & $-47$ &  -89,28  & 238  & 202  & +220,10   & $-1$ &   -3,29  \\ 
N & $R_c$    & 369.8 & 388.4,3.3  & 0 &  1.8,2.1 & $0$  &   +5,25    & $ 356$ & $437$ & +371,57  & -50  & -105 &  -47,22   & $-1$ &   +6,39  \\ 
N & $I_c$    & 303.9 & 343.8,10   & 0 &  2.0,2.5 & $-9$ &  +67,29    & $ 700$ & $684$ & +702,65  & -350 & -357 & -344,19   & $-1$ &  +37,18  \\ 
\hline
\end{tabular}
\end{center}
\end{footnotesize}
\end{table*}

\subsection{Estimation}
\label{subs:IPestim}

From equations (\ref{eq:RobsQ}) and (\ref{eq:RobsU}), it follows that the estimation of IPS can be made by observing objects with $\mathcal{R}_0$ deviating from unity by no more than the precision of the experiment, $10^{-3}$ in our case. Single, non--variable, main--sequence dwarfs appear as good targets in this respect. Although they lack total intrinsic polarization, some polarized flux is generated by scattering on free electrons in atmosphere. The fraction of polarization rises from the centre of the star to the limb and there reaches a value of several per cent for A0 stars. 

$\mathcal{R}_0\ne1$ for these objects because the distribution of polarized flux deviates from the distribution of total flux. However, for A0V stars not closer than 3 pc $|\mathcal{R}_0|-1<10^{-3}$ in the spatial frequency range accessible by the 2.5-m telescope at $\lambda=0.5$~$\mu$m. For the main--sequence stars of later types, the polarization is even smaller.  

For calibration targets, we chose stars closer than 160 pc to ensure a low level of interstellar polarization and brighter than $V<7$~mag to have a good SNR in $\mathcal{R}$. In Table \ref{table:nonpol}, we give the parameters of the calibration stars used in this work. 



The observations of unpolarized stars at the Cassegrain focus are given in Table \ref{table:observationsNonpol}; in total, 13 observations were made. As long as the instrumental effects are expected to be related to the telescope, the polarimetry is given in the horizontal reference system. Total polarization of these stars is $\lesssim10^{-4}$, except for the HIP109121, for it the polarization fraction is $\approx10^{-3}$. This can be attributed to small interstellar polarization. 

We investigated whether the observed $\mathcal{R}$ significantly deviates from unity. For this we calculated $\chi_r^2$ statistics (equation (\ref{eq:chir2def})), which we will denote as $\chi_{r,0}^2$ for this specific case ($\mathcal{R}=1$ hypothesis). For the results see Table \ref{table:observationsNonpol}. For the 13 observations of unpolarized stars, $\chi_{r,0}^2<1.7$ with an average equal to 1.2. Observations of these stars are consistent with the hypothesis that $\mathcal{R}=1$. Observations of HIP20156 secured on 2017 March 7 has $\chi_{r,0}^2=3.1$, which is acceptable.


Although we did not find a deviation of $\mathcal{R}$ from unity for individual unpolarized stars, the existence of a small amount of IPS is not excluded at this stage. To estimate the IPS more reliably we performed fitting of $\mathrm{arg}\mathcal{R}$ by the plane and obtained polaroastrometric signal $s_q^{\star},t_q^{\star},s_u^{\star},t_u^{\star}$ for each observation, with the results presented in Table \ref{table:observationsNonpol2}. These values were averaged for 7 observations conducted through the $V$ filter, see Table \ref{table:IPAS}. One can consider these values to be estimations of IPS. All components of IPS are $<15\,\mu$as, and they do not deviate from zero significantly. In that table, the averaged dimensionless Stokes parameters in a horizontal reference system are given as well. These values have meaning of instrumental polarization, which is less than $10^{-4}$ in our case.

For the estimation of IPS at the Nasmyth focus, we observed 7 unpolarized stars in all filters; see Table \ref{table:observationsNonpol} and \ref{table:observationsNonpol2} for results. 
For each observation, we computed the effective wavelength and placed the corresponding points in Figure \ref{fig:instrPAsignal}. As one can see, the agreement between observed and expected IPS is good, especially for medium--band filters. This agreement is achieved without any direct approximation of polaroastrometric measurements.

\begin{table*}
\begin{center}
\caption{Comparison of observations of $\mu$~Cep and RY~Tau obtained at the Cassegrain and Nasmyth foci. UT and parallactic angle are given at the exposure centre. Dimensionless Stokes parameters and $\mathcal{R}$ obtained at the Nasmyth focus were corrected for instrumental polarization effect. $\chi_{r,0}^2$ is computed under the assumption that $\mathcal{R}$ corrected for instrumental polarization effect equals 1. $\chi_{r,\Delta}^2$ is computed under the assumption that difference in corrected $\mathcal{R}$ between the Cassegrain and Nasmyth foci is zero. $\sigma_0$ is the rms fluctuation of $\mathcal{R}$ and $\sigma_\Delta$ is the rms fluctuation of difference in $\mathcal{R}$. $f_\mathrm{low}$ and $f_\mathrm{high}$ are borders of frequency domain where the $\sigma$ and $\chi_r^2$ values were computed. They are expressed in terms of the cut--off frequency $f_c$. \label{table:correction}}
\begin{footnotesize}
\tabcolsep=0.14cm
\newcolumntype{d}[1]{D{,}{\pm}{#1}}
\begin{tabular}{cccccd{3}d{3}cccccc}
\hline
object & filter & focus & UT                     & $\psi$ & \multicolumn{1}{c}{$q\times10^4$} & \multicolumn{1}{c}{$u\times10^4$} & $\sigma_{0}$& $\chi_{r,0}^2$ & $\sigma_{\Delta}$ & $\chi_{r,\Delta}^2$ & $f_\mathrm{low}/f_c$ & $f_\mathrm{high}/f_c$\\
     &             & &           & $^{\circ}$         &            &           &        &      &        &    & &  \\ 
\hline
$\mu$ Cep & 550 & C     & 2017-12-02T18:22:56 & 81.8  & -24.6,4.2  & -3.8,4.2  & 0.0101   & 17.6 &        &       & &\\ 
          &     & N     & 2017-12-03T16:12:21 & 112.3 & -9.7,17.0   & 1.1,17.0   & 0.0129   & 13.0 & 0.0073 & 1.6    & 0.05 & 0.5\\ 
$\mu$ Cep & 625 & C     & 2017-12-02T18:28:23 & 80.8  & -24.9,5.0  & 8.6,5.0   & 0.0101   & 23.8 &        &        \\ 
          &     & N     & 2017-12-03T16:17:33 & 110.7 & -21.1,16.9  & 1.5,16.9   & 0.0109   & 12.9 & 0.0063 & 1.5     & 0.05 & 0.6\\ 
$\mu$ Cep & 880 & C     & 2017-12-02T18:33:49 & 79.8  & -25.2,5.8  & 39.1,5.8  & 0.0073   & 35.0 &        &        & & \\
          &     & N     & 2017-12-03T16:26:57 & 107.9 & -25.0,19.1 & 11.5,19.3 & 0.0082   & 47.6 & 0.0035 & 1.8     & 0.05 & 0.7\\
RY Tau    & $R$ & N     & 2017-11-03T20:51:39 & -46.7 & 260.8,16.2 & 339.4,16.1 & 0.0545 & 65.7 & & & & \\ 
          &     & N     & 2017-11-04T00:02:16 & 42.9 & 235.3,17.1 & 326.2,17.0 &0.0607 & 67.5 &0.0168 & 1.0 & 0.04 & 0.30 \\           
RY Tau    & $I$ & N     & 2017-09-05T22:46:02 & -55.3 & 276.3,15.1 & 426.5,15.1 & 0.0684 & 97.4 & & & & \\ 
          &     & N     & 2017-09-06T00:42:01 & -47.1 & 274.5,15.1 & 426.1,15.1 &0.0678 & 119.3 &0.0133 & 0.7 & 0.05 & 0.40 \\ 
          \hline
\end{tabular}
\end{footnotesize}
\end{center}
\end{table*}

It is of mild concern that measured components $s^{\star}_{q,\mathrm{ins}}$ and $t^{\star}_{q,\mathrm{ins}}$ of IPS deviate from zero significantly and reach $\approx70~\mu$as for the $I_c$ filter, while the expected level is smaller than $10~\mu$as. 

The IPS expected from the model of instrumental polarization is given in Table \ref{table:IPAS}. For the components $t_q^{\star}$ and $s_u^{\star}$, which depend most on the wavelength we calculated the expected IPS using typical spectra of A0V and M2V stars \citep{Gunn1983}. 

For the $V, R_c$, and $I_c$ filters, the difference in IPS between white and red stars can be as large as 80~$\mu$as, which corresponds to a variation in $\mathrm{arg}\mathcal{R}$ of $1.1\times10^{-2}$ at $0.5f_c$. If one wants to achieve better $\mathcal{R}$ precision, the IPS should be measured on objects with spectra similar to the scientific object. At the same time, in the medium band filters 550, 625 and 880, the dependence of IPS on spectra is $<15\,\mu$as, which is smaller than the precision of IPS determination. Thus, A--F stars can be used as calibration sources for any objects in these filters.

We adopted the averaged observations for the final values of IPS and the spread of observations for the error of determination of the latter, see Table \ref{table:IPAS}. 

\subsection{Correction}

We based the correction of DSP measurements for the instrumental polarization effect on equations (\ref{eq:RobsQ}) and (\ref{eq:RobsU}). First, the measured $\mathcal{R}_Q$ and $\mathcal{R}_U$ were converted to the horizontal reference system (Appendix \ref{app:rotation}). Then, they were divided by $\mathcal{R}_{Q,\mathrm{ins}}$ and $\mathcal{R}_{U,\mathrm{ins}}$, which were in turn computed using equations (\ref{eq:insSlope1}) and (\ref{eq:insSlope2}). The result was converted back to the equatorial reference system.

For the test of this procedure, we observed $\mu$~Cep at the Cassegrain and Nasmyth foci in filters 550, 625, and 880 at epochs spaced by $\approx1^\mathrm{d}$. In addition, a young stellar object RY~Tau was observed at the Nasmyth focus in $R_c$ and $I_c$ filters at close epochs but at different parallactic angles. Details of these observations can be found in the electronic version of Tables \ref{table:observationsNonpol} and \ref{table:observationsNonpol2}. The observations secured at the Nasmyth focus were corrected for instrumental effect as described above. 

$\mu$~Cep demonstrates variability in both total flux and polarization. The timescale of this variability is $\gtrsim100^\mathrm{d}$ \citep{Polyakova2003}. It is unlikely that the object significantly changed its appearance between the considered epochs. For an additional check, one can see from Table \ref{table:correction} that polarimetry for the considered epochs lies within the error of measurement. 

Although RY~Tau can significantly change its brightness and polarization on timescales of several hours \citep{Oudmaijer2001}, sometimes it is stable. For the test of the correction procedure we chose the periods of stability, as evidenced by Table \ref{table:correction}. The $\mathcal{R}$ value can be compared directly for these epochs.

\begin{figure}
\begin{center}
\includegraphics[width=8cm]{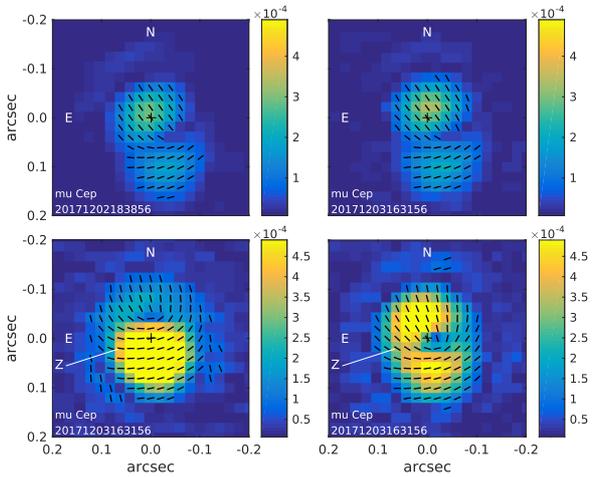}
\end{center}
\caption{Demonstration of the correction for instrumental effect in reconstructed polarization images. Upper left: Cassegrain focus, no correction needed. Lower left: Nasmyth focus, no correction applied. Lower right: Nasmyth focus, only correction for instrumental polarization applied (assume that IPS is zero). Upper right: Nasmyth focus, full correction is applied. Direction to zenith is indicated by Z. 
\label{fig:IPScorrDemo}}
\end{figure}



The comparison was performed by computation of $\chi_r^2$ for the hypothesis that the difference between $\mathcal{R}$ values in two considered epochs should be zero. The result is presented in Table \ref{table:correction} as $\chi_{r,\Delta}^2$. One can see that the difference does not significantly deviate from zero. In Table \ref{table:correction} we give the rms of absolute and differential $\mathcal{R}$ as well. The latter is several times smaller than the former. For $R_c$ and $I_c$ filters it is smaller than $1.7\times10^{-2}$; for 550, 625 and 880~nm filters it is smaller than $0.8\times10^{-2}$. We adopt these numbers as typical of the precision of correction.

Another way to demonstrate the validity of correction is by visual comparison of images reconstructed using the method from Section \ref{sec:recon}; see Figure \ref{fig:IPScorrDemo}. One can see that polarimetric images computed from $\mathcal{R}$ obtained at the Nasmyth focus without correction for instrumental effect are very different from those obtained at the Cassegrain focus. They are dominated by the leak of the polarized flux of the central bright star induced by instrumental polarization of the telescope and by differential polarimetric aberrations. On the other hand, the image in polarized flux after the correction complies reasonably well with that obtained at the Cassegrain focus. All major details of the $\mu$~Cep envelope are reproduced. Nevertheless, the Nasmyth image appears shifted in the North direction by 20~mas with respect to the Cassegrain image. This shift can be attributed to non--ideal correction for the instrumental effect.

The correction of polaroastrometry for IPS is done in a similar way. First, the signal is converted to the horizontal reference system, then the IPS is subtracted, and finally, the conversion to the equatorial reference system is applied. The precision of correction can be evaluated from the numbers presented in Table \ref{table:IPAS}.

%



\section{Discussion}
\label{sec:discussion}

To compare the performance of speckle interferometry and DSP, we took the series for HIP109121 and $\mu$~Cep and computed the average power spectrum in one of the beams of the polarimeter. The result is presented in the left panels of Figure \ref{fig:RvalPolnonpolCompar}. One can see that the power spectra are affected by residual atmospheric and Wollaston prism dispersion. One can hardly tell that there is any departure from a point source in the case of $\mu$~Cep. 

Thanks to employment of differential polarimetric measurement, both the random and systematic components of error of observable $\mathcal{R}$ are much smaller than in the case of a simple averaged power spectrum. The faint signal of $\mu$~Cep is clearly detected. Note that it has an amplitude as small as 0.01.

\begin{figure}
\begin{center}
\includegraphics[width=8cm]{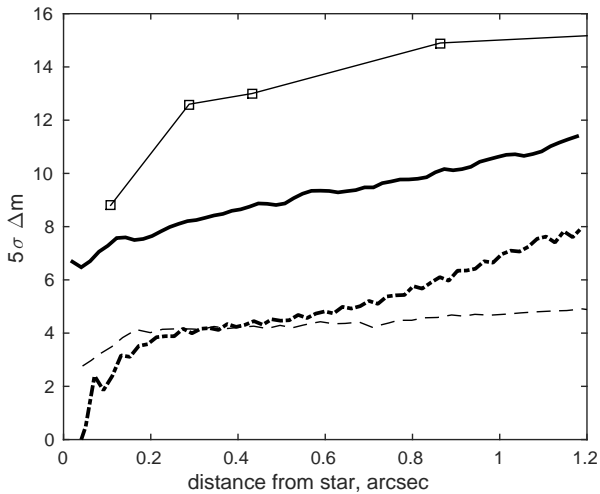}
\end{center}
\caption{Typical detection threshold of a faint point source. The thick dash--dotted line stands for the total flux of the source, which can be detected by conventional speckle interferometry. The thick solid line stands for the polarized flux of the source which can be detected by DSP. In both cases, the same data is used: HIP71251, $R_c$ band, magnitude $R=6.2$, a total accumulation time 220~s, and a total number of accumulated photons of $1.5\times10^9$. The thin dashed line stands for typical detection threshold for the SOAR speckle interferometer \citep{Tokovinin2010} (total flux). The thin line with squares stands for typical detection threshold for the VLT/ZIMPOL \citep{Schmid2018} (polarized flux).
\label{fig:contrast}}
\end{figure}

The speckle interferometry and DSP can be compared in terms of their capability to detect faint point--like sources as well. For the former method the task is to detect faint secondary peaks in the autocorrelation function (ACF). In case of DSP, one attempts to do so in the polarized intensity image reconstructed using the method described in Section \ref{sec:recon}. Remember that this reconstruction is fairly precise for the faint envelopes. We compared these approaches using the series for the unpolarized star HIP71251 obtained on 2018 May 27 in the $R_c$ band. 

For the estimation of the detection limit in the case of speckle interferometry we used the methodology of \citet{Tokovinin2010}. The ACF was computed from the power spectrum. Then, the rms $\sigma$ of ACF fluctuations in the rings with a width of 3 pixels was computed as a function of the angular distance to the origin. We adopted the product of $5\sigma$ and the fraction of PSF energy contained in a $3\times3$ square (0.58) as the detection threshold. This value, expressed as the difference in magnitudes between bright and faint components, is plotted in Figure \ref{fig:contrast}.

\begin{figure}
\begin{center}
\includegraphics[width=5cm]{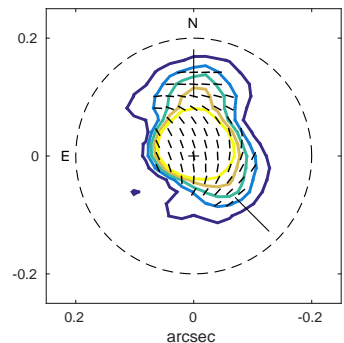}
\end{center}
\caption{Distribution of the polarized intensity for RY Tau. Contours correspond to the ratio of polarized flux in pixel to total flux of the object: $(1.5, 3.0, 4.5, 6.0, 7.5)\times10^{-4}$. The detector was binned $2\times2$, thus the angular scale is $0.0412$~arcsec. The short lines indicate polarization orientation. The two longer solid lines are for directions PA $\approx0^{\circ}$ and PA $\approx225^{\circ}$ indicating two detected polarized features. The dashed circle corresponds to the region covered by the coronagraphic mask in \citep{Takami2013}.
\label{fig:RYTau_img}}
\end{figure}

A contrast of $4$~mag at distances of $0.25$~arcsec is achieved, which is typical for speckle interferometers \citep{Tokovinin2010,Horch2011}. The detection limit deteriorates slowly with decreasing distance to the star until a knee is reached at $\approx0.2$~arcsec. At smaller distances, the contrast starts to degrade faster and reaches $0$~mag in the origin.

Polarimetric images restored from $\mathcal{R}$ were processed in a similar way in order to estimate the detection limit for the faint polarized point source; see Figure \ref{fig:contrast}. In polarized light DSP allows to detect sources by $4$~mag fainter than with conventional speckle interferometry. This gain is comparable with the effect of the application of polarimetric differential imaging (PDI) to coronagraphic instruments equipped with adaptive optics \citep{Hinkley2009}. 

In Figure \ref{fig:contrast} we plotted for comparison the detection limit curve in terms of polarized flux for VLT/ZIMPOL \citep{Schmid2018}. This coronagraph operates in the wavelength range from 600~nm to 900~nm. With ZIMPOL, it is possible to detect polarized point sources by $4-5$~mag fainter than with SPP at least at distances $>0.1$~arcsec from the star. 

At distances $\gg\lambda/D$, a coronagraph will always perform better than polarimetric interferometry as long as it separates the photon noise of the star from the signal of faint source much more effectively. At the same time, in a region closer to the star, the necessity for precise calibration of polarization aberrations and the ability to cope with the high flux of the central star make the situation more favourable for polarimetric interferometry and in particular DSP. 

To demonstrate how DSP can augment coronagraphic data, we consider the observation of RY~Tau obtained using SPP in the $I$ band on 2017 Mar 3 (for details see Table \ref{table:observationsNonpol}). In Figure \ref{fig:RYTau_img}, we present the image in polarized intensity. There are two maxima of intensity in the image at PA $\approx0^{\circ}$ and PA $\approx225^{\circ}$ at a stellocentric distance of $<0.15$~arcsec. Their location is in accordance with features of RY~Tau circumstellar environment found by \citet{Takami2013} using Subaru/HiCIAO at distances of $\gtrsim0.2$~arcsec from the star. \citet{Takami2013} did not investigate the regions of circumstellar environment closer to the star as long as the latter was covered by the coronagraphic mask. The image reconstructed by DSP demonstrates that features found by \citet{Takami2013} can be traced down to $\approx0.05$~arcsec from the star. In addition, one can see that a large total polarization of the source is generated in a region smaller than $\approx0.05$~arcsec and is associated with the star.





Although the images produced by applying the inverse Fourier transform are seemingly easy to interpret, they should be handled with caution. One should remember that they are convoluted with OTF, either diffraction limited or truncated at a frequency smaller than $f_c$, which introduces ambiguity in the result \citep{Min2012}. Additionally, the finite angular size of the central star affects the restored Stokes distribution. Brighter envelopes are restored with worse precision; see Section \ref{sec:recon}.

Due to these factors, it is recommended to perform quantitative comparison between observables and models in Fourier space, as usual in interferometry. This is especially relevant for working on angular scales $\approx\lambda/D$ --- a region where DSP is expected to have its niche. 

\section{Conclusion}
\label{sec:conclusion}





Polarimetric interferometry is a new and yet underestimated method allowing the study of polarized features of astrophysical objects at diffraction--limited resolution. Its basic observable is the ratio $\mathcal{R}$ of visibilities of the object in two orthogonal polarizations. We investigate a method for estimation of this quantity from a series of short--exposure images obtained with a dual--beam polarimeter without AO correction --- Differential Speckle Polarimetry (DSP). The study is performed on the basis of a specialized instrument SPeckle Polarimeter (SPP).

Measurements of $\mathcal{R}$ secured at the Nasmyth focus are influenced by the instrumental polarization effect. This influence can be considered as a simple bias. It can be calibrated out using measurements of unpolarized stars. The precision of correction is $1.8\times10^{-2}$ for observations in Bessel filters and two times smaller for medium--band filters. For more precise measurement it is recommended to use the Cassegrain focus. There the typical precision of $\mathcal{R}$ amplitude and phase determination is $5\times10^{-3}$ for objects with magnitude $I=6$ and can be several times smaller for brighter objects.

The $\mathcal{R}$ value can be used for restoration of polarized intensity image of the faint envelope around a bright star. Contrast of $8$~mag is accessible at a stellocentric distance of $0.25$~arcsec for $I=6$ stars. As long as the instrument does not incorporate a coronagraphic mask it is possible to obtain information on the polarized flux distribution at distances of $\approx\lambda/D$ from the star. Sources of polarized flux even closer to the star can be characterized in terms of the polaroastrometric signal --- the displacement of the object photocentre in orthogonal polarizations. The precision of such measurements at the Cassegrain focus amounts to $15-20~\mu$as for $I=6$ stars.

DSP has potential as an inexpensive way to study polarized circumstellar envelopes at distances of $\approx\lambda/D$ from central star. In this region of stellocentric distances, the envelopes are bright due to higher density and proximity to the star so that the required contrast can be achieved without adaptive optics and coronagraph. This is especially relevant as long as the application of the latter methods in the region $\approx\lambda/D$ is complicated. The protoplanetary disks and dusty environment of evolved stars are the examples of objects for which the DSP is suitable. Some objects of these types are currently under study with the SPP. The results will be published in subsequent papers. 

The next step in the development of the DSP methodology is reaching consistency between the observed level of $\mathcal{R}$ value noise and the results of the simulation presented previously by \citet{Safonov2013}. Thus, we will identify main factors degrading the SNR and determine ways to upgrade the instrument and methodology. The expected gain from the implementation of DSP after AO, either low--order or extreme, will be estimated as well.

\section*{Acknowledgements}
We thank the staff at the Caucasian Observatory of SAI MSU for the help with construction and operation of the SPP. This work has been supported by the Russian Foundation of Basic Research, project no. 16-32-60065. The development and construction of the speckle polarimeter of 2.5-m telescope has been funded by the M.V.Lomonosov Moscow State University Program of Development.



\bibliography{SPPinstrument2b}   
\bibliographystyle{mnras}


\appendix

\section{Average $\mathcal{R}$}
\label{app:averR}

In our paper \citep{Safonov2013} we considered the images in a dual--beam polarimeter obtained through the turbulent atmosphere. Now, we add a rotating HWP. As in the original paper, we assume that the total polarized flux from the object is small in comparison with total flux. Instrumental polarization is assumed to be small as well. We find a first--order approximation of $\mathcal{R}$.

The Fourier transforms of the left and right images will be:
\begin{equation}
F_L = 0.5 \bigl(S_I + S_Q \cos(4\theta) + S_U \sin(4\theta)\bigr),
\end{equation}
\begin{equation}
F_R = 0.5 \bigl(S_I - S_Q \cos(4\theta) - S_U \sin(4\theta)\bigr),
\end{equation}
where $\theta$ is position angle of HWP. The $S_I$, $S_Q$ and $S_U$ have the following form:
\begin{equation}
S_I = (T_{AA^*}+T_{DD^*})I+(T_{AA^*}-T_{DD^*})Q,
\end{equation}
\begin{equation}
S_Q = (T_{AA^*}-T_{DD^*})I+(T_{AA^*}+T_{DD^*})Q,
\end{equation}
\begin{equation}
S_U = (T_{AC^*}+T_{CA^*}+T_{BD^*}+T_{DB^*})I+(T_{AD^*}+T_{DA^*})U.
\end{equation}
The $T_{XY^*}$ terms are expressed as:
\begin{equation}
T_{XY^*}(\boldsymbol{f}) = \int P_X(\boldsymbol{x}) P_Y^*(\boldsymbol{x}+\lambda\boldsymbol{f}) \exp\bigl\{-i\bigl(\phi(\boldsymbol{x})-\phi(\boldsymbol{x}-\lambda\boldsymbol{f})\bigr)\bigr\}d\boldsymbol{x}.
\end{equation}
$\lambda$ is the wavelength and $\boldsymbol{x}$ is a two--dimensional coordinate in pupil plane. The integration takes place over the whole pupil space. $P_X(\boldsymbol{x})$ are the components of the so--called Jones pupil \citep{Breckinridge2015}. $\phi$ is the instantaneous phase disturbed by the atmosphere.

Let us substitute the $F_L$ and $F_R$ into Equations (\ref{eq:RavgC}) and (\ref{eq:RavgS}) at $h=4$:
\begin{equation}
\mathcal{R}_{c4} = 1 + \frac{2\langle S_Q S_I^*\rangle}{\langle S_I S_I^*\rangle},\,\,\,\mathcal{R}_{s4} = 1 + \frac{2\langle S_U S_I^*\rangle}{\langle S_I S_I^*\rangle}.
\end{equation}

As the next step, we introduce the quantities $\Delta T_{Q1} = T_{AA^*}-T_{DD^*}$ and $\Delta T_{U1} = T_{AC^*}+T_{CA^*}+T_{BD^*}+T_{DB^*}$.  These quantities are small relative to conventional instantaneous OTF $T_\mathrm{atm}$. Retaining the first--order approximation it is possible to state that:
\begin{equation}
\mathcal{R}_{c4} \approx 1 + \frac{2\langle T_\mathrm{atm}^* \Delta T_{Q1} \rangle}{\langle T_\mathrm{atm} T_\mathrm{atm}^* \rangle} + \frac{2Q}{I},
\label{eq:TatmInstrQ}
\end{equation}
\begin{equation}
\mathcal{R}_{s4} \approx 1 + \frac{2\langle T_\mathrm{atm}^* \Delta T_{U1} \rangle}{\langle T_\mathrm{atm} T_\mathrm{atm}^* \rangle} + \frac{2U}{I}.
\label{eq:TatmInstrU}
\end{equation}
In \citep{Safonov2013}, we demonstrated that the second terms in these equations converge to values which depend only on the optical scheme of the telescope. Therefore, they constitute instrumental bias. 

Equations (\ref{eq:TatmInstrQ}) and (\ref{eq:TatmInstrU}) can be transformed in order to yield (\ref{eq:RobsQ}) and (\ref{eq:RobsU}).

\section{Rotation and error of $\mathcal{R}$}
\label{app:rotation}

Under the assumption that the total polarized flux is small relatively to the total flux, it is possible to transform $\mathcal{R}_Q$ and $\mathcal{R}_U$ from one reference system to another, rotated by $\psi$ counter--clockwise. Let us introduce auxiliary notations: $\Delta\mathcal{R}_Q = \mathcal{R}_Q-1$ and $\Delta\mathcal{R}_U = \mathcal{R}_U-1$. Thus, $\mathcal{R}$ value in new reference system will be:
\begin{equation}
\Delta\mathcal{R}_Q^\prime(f_x^\prime,f_y^\prime) = \cos2\psi\Delta\mathcal{R}_Q(f_x,f_y) + \sin2\psi\Delta\mathcal{R}_U(f_x,f_y),
\end{equation}
\begin{equation}
\Delta\mathcal{R}_U^\prime(f_x^\prime,f_y^\prime) = -\sin2\psi\Delta\mathcal{R}_Q(f_x,f_y) + \cos2\psi\Delta\mathcal{R}_U(f_x,f_y),
\end{equation}
where
\begin{equation}
f_x = \cos\psi f_x^\prime - \sin\psi f_y^\prime,
\end{equation}
\begin{equation}
f_y = - \sin\psi f_x^\prime + \cos\psi f_y^\prime,
\end{equation}

While $\mathcal{R}_{c4}$ and $\mathcal{R}_{s4}$ contain useful signal, the other $\mathcal{R}_{ch}$ and $\mathcal{R}_{sh}$ ($h\ne4$) should be zero. In practice, they are not, due to noise. One can assume that this noise affects all harmonics equally. Therefore, $\mathcal{R}_{ch}$ and $\mathcal{R}_{sh}$ ($h\ne4$) can be used for evaluation of noise and the subsequent estimation of $\mathcal{R}_{c4}$, $\mathcal{R}_{s4}$ errors:
\begin{equation}
\sigma_{\mathcal{R},\mathrm{abs}}^2(\boldsymbol{f}) = \frac{1}{N_h-4} \sum_{h=5}^{N_h} (|\mathcal{R}(\boldsymbol{f})|-1)^2,
\end{equation}
\begin{equation}
\sigma_{\mathcal{R},\mathrm{arg}}^2(\boldsymbol{f}) = \frac{1}{N_h-4} \sum_{h=5}^{N_h} \mathrm{arg}\mathcal{R}^2(\boldsymbol{f}).
\end{equation}
These relations can be formulated for Stokes $Q$ and $U$. A similar procedure was proposed by \citet{Bagnulo2009} for polarimetry. 

Using error estimation, one can devise a metric characterizing goodness of fit of the observed $\mathcal{R}_\mathrm{obs}$ by some model $\mathcal{R}_\mathrm{mod}$:
\begin{equation}
\begin{split}
\chi_r^2 & = \frac{1}{4N} \sum_{\boldsymbol{f}} \frac{\bigl(|\mathcal{R}_{Q,\mathrm{obs}}(\boldsymbol{f})|-|\mathcal{R}_{Q,\mathrm{mod}}(\boldsymbol{f})|\bigr)^2}{\sigma_{\mathcal{R},Q,\mathrm{abs}}^2(\boldsymbol{f})} \\
         & + \frac{1}{4N} \sum_{\boldsymbol{f}} \frac{\bigl(\mathrm{arg}\mathcal{R}_{Q,\mathrm{obs}}(\boldsymbol{f})-\mathrm{arg}\mathcal{R}_{Q,\mathrm{mod}}(\boldsymbol{f})\bigr)^2}{\sigma_{\mathcal{R},Q,\mathrm{arg}}^2(\boldsymbol{f})} \\
         & + \frac{1}{4N} \sum_{\boldsymbol{f}} \frac{\bigl(|\mathcal{R}_{U,\mathrm{obs}}(\boldsymbol{f})|-|\mathcal{R}_{U,\mathrm{mod}}(\boldsymbol{f})|\bigr)^2}{\sigma_{\mathcal{R},U,\mathrm{abs}}^2(\boldsymbol{f})} \\
         & + \frac{1}{4N} \sum_{\boldsymbol{f}} \frac{\bigl(\mathrm{arg}\mathcal{R}_{U,\mathrm{obs}}(\boldsymbol{f})-\mathrm{arg}\mathcal{R}_{U,\mathrm{mod}}(\boldsymbol{f})\bigr)^2}{\sigma_{\mathcal{R},U,\mathrm{arg}}^2(\boldsymbol{f})}. \\
\end{split}
\label{eq:chir2def}
\end{equation}
Here, the summation is performed over some region of Fourier space. $N$ is the number of observational points in Fourier space falling in that region.

\bsp	
\label{lastpage}
\end{document}